\documentclass[aps,prl,nofootinbib,superscriptaddress,preprintnumbers,amsmath,amssymb,latexsym,array,enumerate,letter,twocolumn]{revtex4}
\pdfoutput=1

\usepackage{times}
\usepackage{latexsym}
\usepackage{graphicx, graphics, hyperref, amsmath, amssymb, slashed, xcolor, bbm,bm,amsthm, array}
 \usepackage{subfigure}
 \usepackage{listings} 
 \usepackage{trimclip}

\usepackage{rotating}
\usepackage{afterpage}

\newcommand{\nc}{\newcommand}

\nc{\beq}{\begin{equation}}
\nc{\eeq}{\end{equation}}
\nc{\barray}{\begin{eqnarray}}
\nc{\earray}{\end{eqnarray}}
\nc{\barrayn}{\begin{eqnarray*}}
\nc{\earrayn}{\end{eqnarray*}}
\nc{\bcenter}{\begin{center}}
\nc{\ecenter}{\end{center}}
\nc{\mc}{\mathcal}
\nc{\er}[1]{(\ref{eq:#1})}
\nc{\onehalf}{\frac{1}{2}} 
\nc{\partialbar}{\bar{\partial}}
\nc{\psit}{\widetilde{\psi}}
\nc{\Tr}{\mbox{Tr}}
\nc{\hc}{\mbox{H.c.}}
\nc{\ev}{\;\mathrm{eV}}
\nc{\mev}{\;\mathrm{MeV}}
\nc{\gev}{\;\mathrm{GeV}}
\nc{\kev}{\;\mathrm{keV}}
\nc{\tev}{\;\mathrm{TeV}}

\def\chii0{\chi_i^0}
\def\chij0{\chi_j^0}

\newcommand{\gsim}{\lower.7ex\hbox{$\;\stackrel{\textstyle>}{\sim}\;$}}
\newcommand{\lsim}{\lower.7ex\hbox{$\;\stackrel{\textstyle<}{\sim}\;$}}
\nc{\ttbar}{t\bar t}

\def\beq{\begin{equation}}
\def\eeq{\end{equation}}
\def\bea{\begin{eqnarray}}
\def\eea{\end{eqnarray}}

\newcommand{\cref}[1]{Chapter~\ref{c.#1}}

\def\beq{\begin{equation}}
\def\eeq{\end{equation}}
\def\bea{\begin{eqnarray}}
\def\eea{\end{eqnarray}}

\graphicspath{{plots/}}

\begin{document}

\widetext

\title{Determining the Neutrino Lifetime from Cosmology}

\author{Zackaria Chacko}
\affiliation{Maryland Center for Fundamental Physics, Department of Physics, University of Maryland, College Park, MD 20742-4111 USA}

\author{Abhish Dev}
\affiliation{Maryland Center for Fundamental Physics, Department of Physics, University of Maryland, College Park, MD 20742-4111 USA}

\author{Peizhi Du}
\affiliation{C.N. Yang Institute for Theoretical Physics, Stony Brook University, Stony Brook, NY, 11794, USA}

\author{Vivian Poulin}
\affiliation{Laboratoire Univers \& Particules de Montpellier (LUPM), CNRS \& Universit\'e de Montpellier (UMR-5299),Place Eug\`ene Bataillon, F-34095 Montpellier Cedex 05, France}

\author{Yuhsin Tsai}
\affiliation{Maryland Center for Fundamental Physics, Department of Physics, University of Maryland, College Park, MD 20742-4111 USA}

\preprint{YITP-SB-20-01, LUPM:20-006}
 \begin{abstract}

 We explore the cosmological signals of theories in which the neutrinos 
decay into invisible dark radiation after becoming non-relativistic. We 
show that in this scenario, near-future large scale structure 
measurements from the Euclid satellite, when combined with cosmic 
microwave background data from Planck, may allow an independent 
determination of both the lifetime of the neutrinos and the sum of their 
masses. These parameters can be independently determined because the 
Euclid data will cover a range of redshifts, allowing the growth of 
structure over time to be tracked. If neutrinos are stable on 
cosmological timescales, these observations can improve the lower limit 
on the neutrino lifetime by seven orders of magnitude, from 
$\mathcal{O}(10)$ years to $2\times 10^8$ years ($95\%$ C.L.), without 
significantly affecting the measurement of neutrino mass. On the other 
hand, if neutrinos decay after becoming non-relativistic but on 
timescales less than $\mathcal{O}(100)$ million years, these 
observations may allow, not just the first measurement of the sum of 
neutrino masses, but also the determination of the neutrino lifetime 
from cosmology.

 \end{abstract}
 
\maketitle


\section{Introduction}

Neutrino decay is a characteristic feature of models in which neutrinos 
have masses. Even in the minimal extension of the Standard Model (SM) 
that incorporates Majorona neutrino masses through the 
non-renormalizable Weinberg operator, the heavier neutrinos are 
unstable, and undergo decay at one loop into a lighter neutrino and a 
photon. The same is true of the minimal extension of the SM that 
incorporates Dirac neutrino masses through the inclusion of right-handed 
singlet neutrinos. In both cases, the lifetime of the heavier neutrino 
is of order $\tau_{\nu} \sim 
10^{50}\textrm{s}\left({0.05\,\textrm{eV}}/{m_{\nu}}\right)^5$, in the 
limit that the daughter neutrino mass is 
neglected~\cite{Petcov:1976ff,Goldman:1977jx,Marciano:1977wx,Lee:1977tib,Pal:1981rm}. 
This is much longer than the age of the universe, and so these minimal 
neutrino mass models do not give rise to observable signals of neutrino 
decay. However, in general, the neutrino lifetime can be much shorter. 
For example, in theories where the generation of neutrino masses is 
associated with the breaking of global 
symmetries~~\cite{Gelmini:1980re,Chikashige_1981,Georgi:1981pg,VALLE198387,Gelmini:1983ea} 
(see also~\cite{Dvali:2016uhn,Funcke:2019grs}), a heavier neutrino can 
decay into a lighter neutrino and a Goldstone boson on timescales that 
can be much shorter than the age of the universe.

Until the turn of the century, the decaying neutrino scenario attracted 
considerable attention as a possible solution to the solar and 
atmospheric neutrino 
problems~\cite{Bahcall:1972my,Berezhiani:1991vk,Acker:1991ej,Barger:1998xk}. 
However, this explanation is now disfavored by the 
data~\cite{Acker:1993sz,CHOUBEY200073,Joshipura:2002fb}. More recently, 
radiative neutrino decays have been put forward as a possible 
explanation of the anomalous 21 cm signal observed by the EDGES 
experiment~\cite{Chianese:2018luo}.

There is a strong lower limit on the neutrino lifetime in the case of 
radiative decays. In this scenario, the limits on spectral distortions 
in the cosmic microwave background (CMB) can be translated into bounds 
on radiative neutrino decays, $\tau_{\nu}\gsim 10^{19}$s for the larger 
mass splitting and $\tau_{\nu}\gsim 4 \times 10^{21}$s for the smaller 
one~\cite{Aalberts:2018obr}, greater than the age of the universe. There 
are also very strong laboratory and astrophysical bounds on the neutrino 
dipole moment operators that induce radiative neutrino 
decays~\cite{Beda:2013mta,Borexino:2017fbd,Raffelt:1990pj,Raffelt:1999gv,Arceo-Diaz:2015pva}.

In contrast, the decay of neutrinos into invisible dark radiation is 
only weakly constrained by current data. At present, the most stringent bounds on invisible neutrino decays are 
from cosmological observations. Although limits have also been placed on 
neutrino decay based on data from Supernova 1987A~\cite{Frieman:1987as}, 
solar neutrinos~\cite{Joshipura:2002fb,Beacom:2002cb,Bandyopadhyay:2002qg}, 
atmospheric neutrinos and long baseline experiments 
~\cite{GonzalezGarcia:2008ru,Gomes:2014yua,Choubey:2018cfz,Aharmim:2018fme}, 
these constraints are in general much weaker. Cosmological measurements 
are sensitive to the neutrino lifetime through the gravitational effects 
of the relic neutrinos left over from the Big Bang, and their decay 
products. If the neutrino lifetime is less than the timescale of 
recombination, then neutrino decay and inverse decay processes are 
active during the CMB epoch. These processes prevent the neutrinos from 
free streaming, leading to observable effects on the heights and 
locations of the CMB peaks~\cite{Peebles,Hu:1995en,Bashinsky:2003tk}. 
Current limits require that the neutrinos be free streaming from 
redshifts $z \gtrsim 8000$ until recombination, $z \approx 
1100$~\cite{Archidiacono:2013dua,Audren:2014lsa,Follin:2015hya,Escudero:2019gfk}
(see also \cite{Kreisch:2019yzn}). This can be translated into a lower 
bound on the neutrino lifetime, $\tau_\nu\geq 4 \times 
10^{8}\,\textrm{s} 
\left({m_\nu}/{0.05\,\textrm{eV}}\right)^3$~\cite{Escudero:2019gfk}, 
much less than the age of the universe. Therefore, at present there is 
no evidence that neutrinos are stable on cosmological timescales, and 
the lifetime of the neutrino remains an open question. 

A knowledge of the neutrino lifetime is of particular importance for the 
determination of neutrino masses from cosmology. At present, the 
strongest upper limit on the sum of neutrino masses, $\sum m_\nu 
\lesssim 0.12$ eV~\cite{Aghanim:2018eyx}, is from cosmological 
observations. However, this bound assumes that the neutrino number 
density and energy distribution have evolved in accordance with the 
standard Big Bang cosmology until the present time. If the neutrinos 
have decayed~\cite{Serpico:2007pt,Serpico:2008zza} or annihilated 
away~\cite{Beacom:2004yd,Farzan:2015pca}, this bound is not valid, and 
must be reconsidered. In particular, in the case of neutrinos that decay 
on cosmological timescales, values of the neutrino masses as large as 
$\sum m_\nu \sim 0.90$ eV are currently allowed by the 
data~\cite{Chacko:2019nej}. 

In the coming decade, major improvements are expected in the precision 
of cosmological observations, which would lead to great advances in 
neutrino physics. The Euclid satellite, scheduled to be launched in 
2022, is expected to measure both the galaxy and the cosmic shear power 
spectra with unprecedented precision, achieving up to sub-percent 
accuracy over the redshift range from $z\sim 
0.5-2$~\cite{Amendola:2012ys}. In the more distant future, the CMB-S4 
experiment \cite{Abazajian:2016yjj} will lead to major advances over 
current CMB observations. This includes improvements in the measurement 
of CMB lensing, which is very sensitive to the neutrino masses. Under 
the assumption that neutrinos are stable, these new measurements will 
allow us to probe values of the neutrino masses smaller than the 
observed neutrino mass splittings and thereby determine $\sum 
m_\nu$~\cite{Archidiacono:2016lnv,Brinckmann:2018owf}. However, if the 
neutrinos are unstable on cosmological timescales, the question of 
whether $\sum m_\nu$ can in fact be determined remains unanswered.

In this paper, we address this question. We consider theories in which 
the neutrinos decay into invisible dark radiation after becoming 
non-relativistic. We show that in this class of models, near-future 
large scale structure (LSS) measurements from Euclid, in combination 
with Planck data, may allow an independent determination of both the 
lifetime of the neutrinos and the sum of their masses. The reason these 
parameters can be independently determined is because Euclid takes 
measurements at multiple redshifts, which allows us to track the growth 
of structure over time. In the case of stable neutrinos, we find that 
these observations will be able to extend the lower bound on the 
lifetime by at least seven orders of magnitude, from $\mathcal{O}(10)$ 
years to $\mathcal{O}(0.1-10)$ Gyrs depending on the neutrino mass, 
without significantly affecting the measurement of the sum of neutrino 
masses. Furthermore, we show that if the neutrinos decay after becoming 
non-relativistic but with a lifetime less than $\mathcal{O}(10^8)$ 
years, these observations may allow the first determination of not just 
the neutrino masses, but also the neutrino lifetime.

\section{Breaking the Degeneracy Between Neutrino Mass and Lifetime}

The sensitivity of cosmological observables to the neutrino masses 
arises from the fact that, after the neutrinos become non-relativistic, 
their contribution to the energy density redshifts like matter, and is 
therefore greater than that of a relativistic species of the same 
abundance. This leads to a faster Hubble expansion, reducing the time 
available for structure formation. The net result is an overall 
suppression of large scale structure~\cite{Bond:1980ha,Hu:1997mj}, (for 
reviews 
see~\cite{Wong:2011ip,Lesgourgues:2018ncw,Tanabashi:2636832,Lattanzi:2017ubx}). 
A larger neutrino mass gives rise to greater suppression, since heavier 
neutrinos become non-relativistic at earlier times, and also contribute 
more to the total energy density after becoming non-relativistic. In the 
case of neutrinos that decay, the extent of the suppression now also 
depends on the neutrino lifetime. The key idea, first discussed 
in~\cite{Serpico:2007pt,Serpico:2008zza}, is that if the neutrinos decay 
into massless species after becoming non-relativistic, the suppression 
in power is reduced. Depending on how late the decay kicks in after the 
neutrinos have become non-relativistic, the magnitude of the suppression 
will vary.

\begin{figure}
 \centering{
 \includegraphics[scale=0.5]{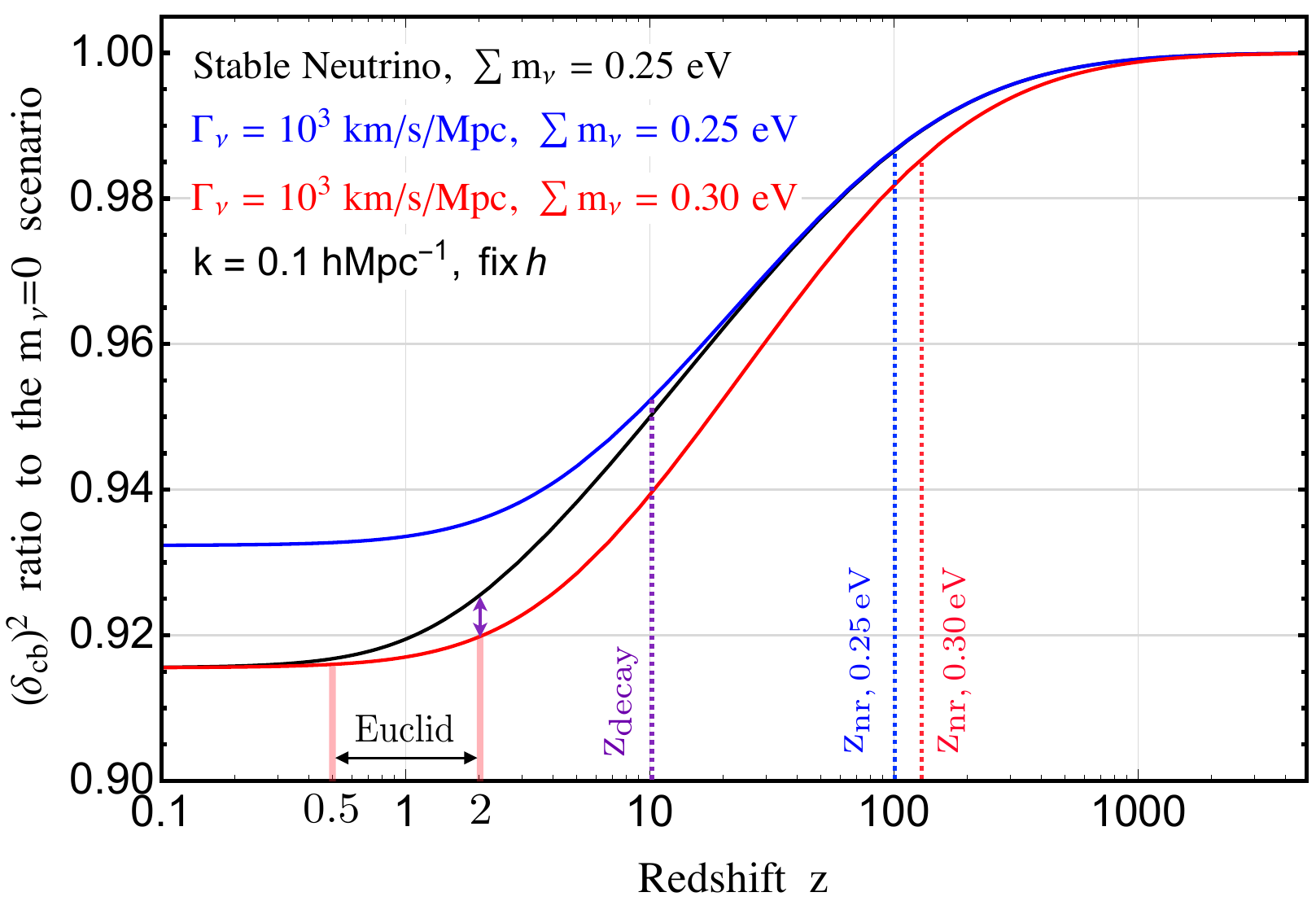}
 }
 \caption{Evolution of the ratio of the CDM+baryon density perturbations 
with respect to the case of massless neutrinos. The blue (black) curve 
corresponds to the case of stable (unstable) massive neutrinos with 
$\sum m_{\nu}=0.25$ eV. Here z$_{{\rm decay}}$, defined as the redshift 
at which the neutrino width $\Gamma_\nu$ becomes equal to the Hubble 
constant, corresponds to the redshift at the time of neutrino decay. 
Similarly z$_{{\rm nr}}$ denotes the redshift at which $80\%$ of the 
neutrinos have become non-relativistic. As compared to stable neutrinos, 
unstable neutrinos generate less suppression of density perturbation at 
low redshift. Unstable heavier neutrinos with $\sum m_{\nu}=0.3$ eV 
(shown in red) can give the same density perturbation at low redshift as 
stable neutrinos of mass $\sum m_{\nu}=0.25$ eV. However, at $z=2$, the 
perturbation in the heavier neutrino scenario deviates at the 
$\mathcal{O}(0.1)\%$ level from the stable neutrino scenario (indicated 
by the purple arrow).}
 \label{fig:zdep}
 \end{figure}

These features are illustrated in Fig.~\ref{fig:zdep}, where we show the 
evolution of the overdensity of cold dark matter and baryons, $\delta_{ 
cb}\equiv\delta\rho_{ cb}/\bar{\rho}_{cb}$, for three cases, based on 
the analysis in~\cite{Chacko:2019nej} and briefly described in the next 
section. The results are expressed in terms of the ratio of 
$(\delta_{cb})^2$ for each case to its value in the scenario with massless 
neutrinos. The black line corresponds to stable neutrinos with $\sum 
m_\nu = 0.25$ eV, while the blue line corresponds to unstable neutrinos 
of the same mass. We see that, as compared to the stable neutrino 
scenario, unstable neutrinos of the same mass lead to a smaller 
suppression of $\delta_{ cb}$ at $z=0$. The red line corresponds to 
unstable neutrinos with $\sum m_\nu = 0.30$ eV, and their lifetime has 
been chosen to obtain the same result for the overdensity at $z=0$ as 
for stable neutrinos with $\sum m_\nu = 0.25$ eV. We see from the black 
and red curves in Fig.~\ref{fig:zdep} that the effects of a stable 
neutrino on the matter density perturbations cannot be easily 
distinguished from those of a heavier neutrino that is shorter-lived 
based only on measurements performed at $z\lsim 0.3$. This is because 
the growth of $\delta_{ cb}$ is almost frozen in the region where the 
cosmological constant dominates ($z\lsim 0.3$). Therefore, there is a 
degeneracy between $\sum m_{\nu}$ and $\tau_{\nu}$ that cannot be 
resolved based only on measurements of the matter power spectrum at low 
redshifts. However, it is clear from Fig.~\ref{fig:zdep} that the 
evolution of the power suppression at earlier times is different in the 
two cases. Consequently, the shapes of the power spectra as a function 
of $z$ are distinct. This would allow these two cases to be 
distinguished if measurements are made at more than one redshift with 
sub-percent precision (e.g., black vs. red at $z=0.5$ and $z=2$ in 
Fig.~\ref{fig:zdep}). As mentioned above, the Euclid experiment is 
expected to take measurements at multiple redshifts between $z \approx 0.5$ 
and $z \approx 2$ at this level of precision. Hence the combined Euclid and 
Planck data has the potential to break the degeneracy between neutrino 
mass and lifetime.

\section{Analysis}

In order to calculate the effects of neutrino decay on cosmological 
observables, we implement the Boltzmann equations corresponding to the 
decay of neutrinos into dark radiation that were derived 
in~\cite{Chacko:2019nej} into the code {\sf 
CLASS}\footnote{http://www.class-code.net}~\cite{Blas:2011rf}. We work 
under the assumption that, after becoming non-relativistic, each SM 
neutrino decays with width $\Gamma_{\nu_i}$ into two massless particles. 
Here the indices $i$ label the neutrino mass eigenstates. For 
concreteness, we assume that the decay widths of the three neutrinos 
satisfy the relation $\Gamma_{\nu_i}\propto m_{\nu_i}^3$. This 
assumption is motivated by models in which the generation of neutrino 
masses is associated with the breaking of global symmetries. In these 
theories the couplings of neutrinos to the Goldstone bosons typically 
scale as $m_{\nu}/f$, where $f$ corresponds to the scale at which the 
global symmetry is broken. Given the observed mass splittings $\Delta 
m_{12}^2=7.4\times 10^{-5}\,\textrm{eV}^2$ and $\Delta 
m_{23}^2=2.5\times 10^{-3}\,\textrm{eV}^2$, this leaves only two 
remaining independent parameters. We choose to present the results of 
our analysis in terms of the parameters $(\sum m_\nu, \Gamma_\nu)$, 
where $\Gamma_\nu$ is the decay width of the heaviest neutrino. With 
this definition, $\Gamma_\nu \equiv \Gamma_{\nu_3}$ for the normal 
hierarchy and $\Gamma_\nu \equiv \Gamma_{\nu_2}$ for the inverted 
hierarchy.

 We wish to determine the extent to which a combination of Planck data 
and future Euclid data can help break the degeneracy between the 
neutrino mass and lifetime. To that end, we make use of the mock 
likelihoods available publicly in {\sc MontePython-v3.1} and described 
in Refs.~\cite{Sprenger:2018tdb,Brinckmann:2018cvx}. We include Euclid 
galaxy and cosmic shear power spectra in the ``realistic'' 
configuration, i.e., we include nonlinear scales and employ a loose 
(redshift-independent) non-linear cut at comoving $k_{\rm NL} = 2~h/{\rm 
Mpc}$ in the galaxy power spectrum and $k_{\rm NL} = 10~h/{\rm Mpc}$ in 
the cosmic shear power spectrum, together with a nonlinear correction 
based on HaloFit \cite{Takahashi:2012em,AliHaimoud:2012vj} and a 
theoretical error on the nonlinear modeling (as described in 
Refs.~\cite{Sprenger:2018tdb,Brinckmann:2018cvx}). For a few cases, we 
employed an alternative ``conservative'' prescription where we cut the 
data at comoving $k_{\rm NL} = 0.2~h/{\rm Mpc}$ in the galaxy power 
spectrum and $k_{\rm NL} = 0.5~h/{\rm Mpc}$ in the cosmic shear power 
spectrum, and verified that this leads to very similar results. This 
gives us confidence in the robustness of our conclusions. In order to 
include {\rm Planck} data in our forecast, we generate a mock dataset 
with the fake likelihood {\sc fake\_planck\_realistic} available in {\sc 
MontePython-v3.1}. We analyze chains using the python package {\sc GetDist} \cite{Lewis:2019xzd}.

\begin{figure*}
\centering
\includegraphics[width=6.5cm]{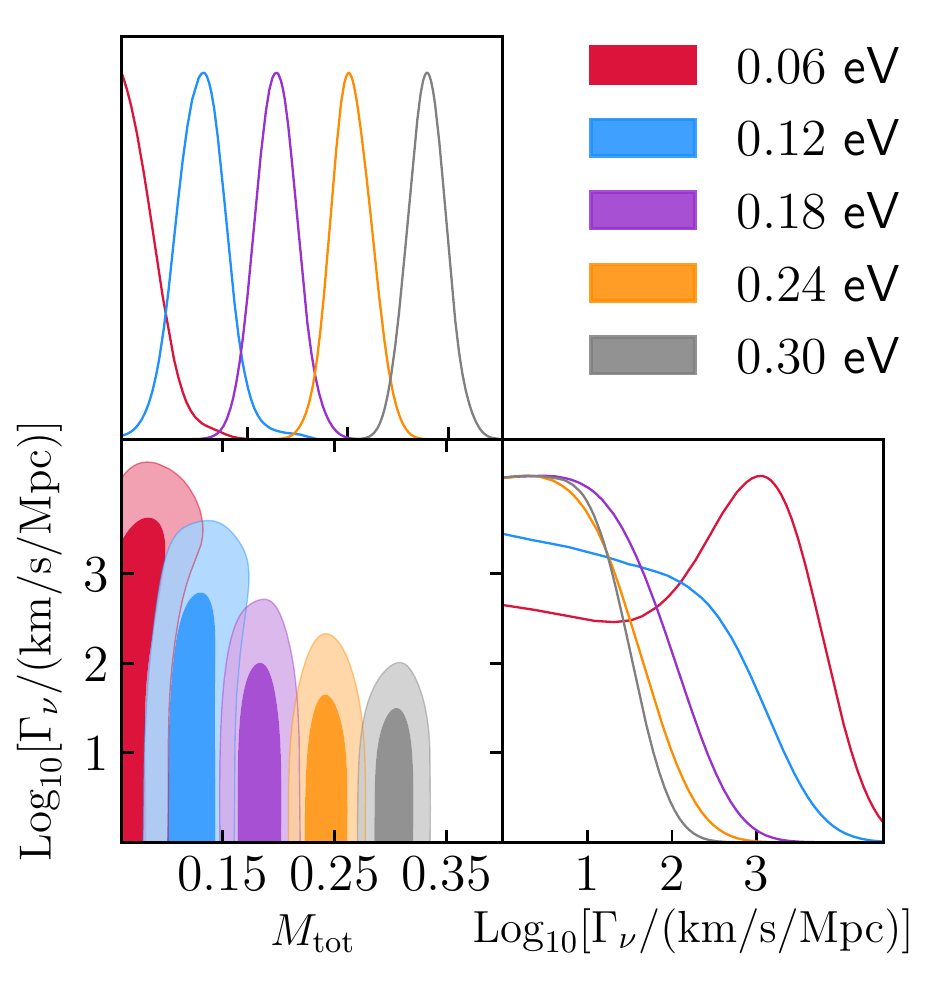}\quad\qquad\includegraphics[width=6.5cm]{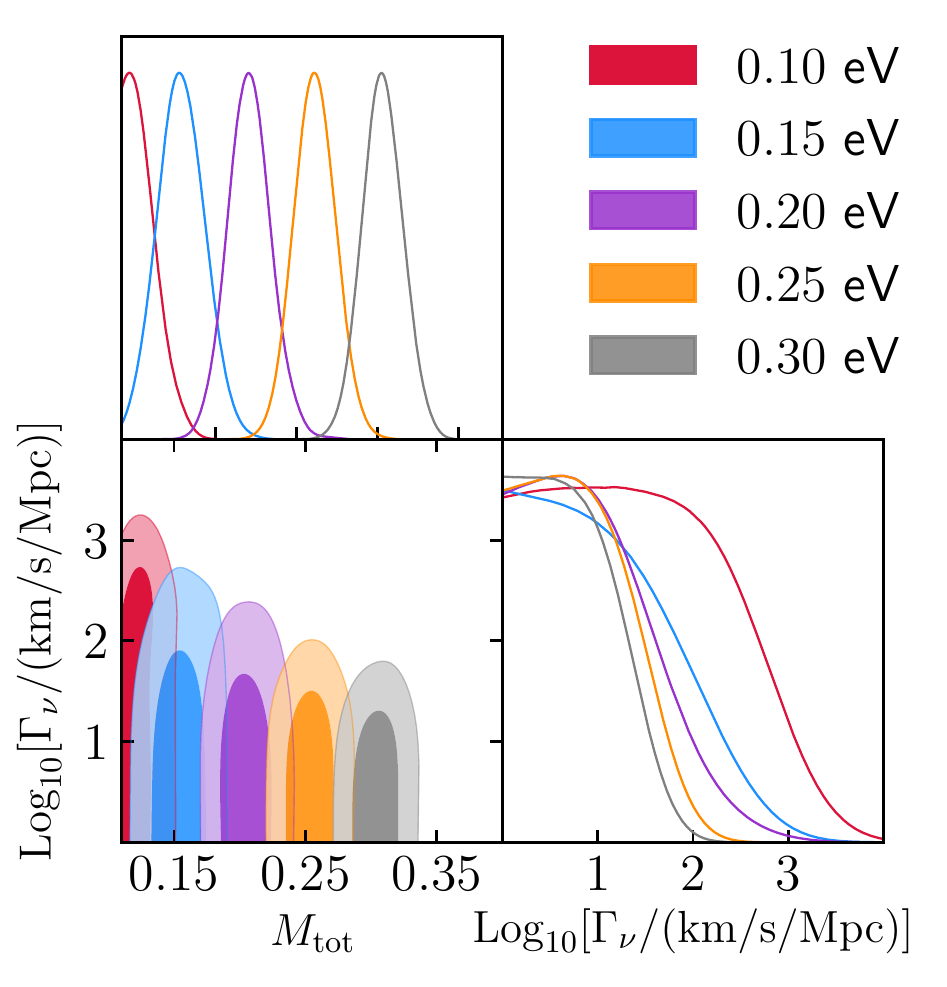}
 \caption{Forecast of the 2D posterior of the sum of neutrino masses and decay width of the heaviest neutrino reconstructed at 68 and 95\% C.L. from a combination of Planck+Euclid $P(k)$+Euclid Lensing. The fiducial model assumes that neutrinos are stable and that they follow the normal hierarchy (left panel) or inverted hierarchy (right panel). } 
 \label{fig:result_stable}
 \centering
 \end{figure*}

We first forecast the lower bound on the neutrino lifetime that can be 
reached in the near future. We begin by generating mock data sets for 
the case of stable neutrinos, i.e., $\Gamma_{\nu}=0$. Specifically, we 
generate a mock data set for the following values of $\sum m_\nu$/eV$: 
[0.06,0.12,0.18,0.24,0.30]$ for the normal hierarchy case and 
$[0.10,0.15,0.20,0.25,0.30]$ for the inverted hierarchy. This range 
covers the minimum $\sum m_\nu$ allowed by the normal and inverted mass 
spectra, and also the maximum $\sum m_\nu$ consistent with the current 
bound derived in~\cite{Chacko:2019nej}. We then run one MCMC scan per 
mock data set varying the $\Lambda$CDM parameters 
$\{\omega_b,\omega_{\rm cdm},100\theta_s,A_s,n_s,\tau_{\rm reio}\}$ 
together with $\{\sum m_\nu/{\rm eV},{\rm Log}_{10}[\Gamma_\nu/({\rm 
km/s/Mpc})]\}$. As mentioned earlier, here $\Gamma_\nu$ refers to the 
width of the heaviest neutrino. As our modifications to {\sc CLASS} have 
the effect of making the code much slower, we are forced to run a large 
number of chains ($\sim100$) to acquire enough points to obtain robust 
results. This penalizes the use of the Gelman-Rubin criterion 
\cite{Gelman:1992zz} as a convergence test{\footnote{Nevertheless, all 
runs satisfy the Gelman-Rubin criterion except for the cases with fiducial $\sum m_\nu/{\rm eV}\!=\!
0.06$,  $\sum m_\nu/{\rm eV}\!=\!0.10$ and  $\sum m_\nu/{\rm eV}\!=\!0.12$. For these runs, we have at most $(R-1) \approx 0.3$, $(R-1) \approx 0.22$ and $(R-1) \approx 0.25$ respectively.}}. Therefore we primarily rely on 
visual inspections, and on comparison between various chunks of chains, 
to assess convergence. As a check, we have verified that for all scenarios, our constraints vary by less than 10\% when 
adapting the fraction of points removed with {\sc GetDist} from 0.1 to 
0.5.

Our results are displayed in Fig.~\ref{fig:result_stable} for the 
normal- (left) and inverted- (right) mass hierarchy cases, where we show 
the bounds on the decay rate $\Gamma_{\nu}$ of the heaviest neutrino as 
a function of $\sum m_\nu$. We summarize the bounds on the neutrino 
masses and lifetime for both hierarchies in Table 
\ref{table:forecast_NH}. Of utmost importance, we find that the 
combination of Planck and Euclid can break the degeneracy between $(\sum 
m_{\nu},\Gamma_{\nu})$ and set an upper bound on the neutrino lifetime, 
${\rm Log}_{10}[\Gamma_\nu/({\rm km/s/Mpc})]\leq 3.7$ ($2\sigma$), even 
for the lowest possible neutrino mass. Moreover, we find that the 
sensitivity to $\sum m_{\nu}$ is not significantly degraded by the 
additional free parameter $\log_{10}\Gamma_{\nu}$. As can be seen from 
Table \ref{table:forecast_NH}, the bounds on $\Gamma_\nu$ in the normal 
and inverted hierarchy cases become increasingly close above $\sum 
m_{\nu}\gsim 0.2$ eV. This is because in this limit the neutrinos are 
becoming quasi-degenerate. Nevertheless, even for $\sum m_{\nu} = 
0.3$ eV, the values of the two largest neutrino masses differ at the 
level of a few percent between the normal and inverted hierarchies. 
Since $\Gamma_{\nu}\propto m_{\nu}^3$, this accounts for the $\sim 10\%$ 
difference between the bounds on $\Gamma_\nu$ in the two cases. Finally, 
we mention that we do not find any strong correlation between the decay 
rate and the other cosmological parameters. Therefore, for brevity we do 
not explicitly report the reconstructed $\Lambda$CDM parameters.

\begin{figure*}
    \centering
    \includegraphics[width=6.5cm]{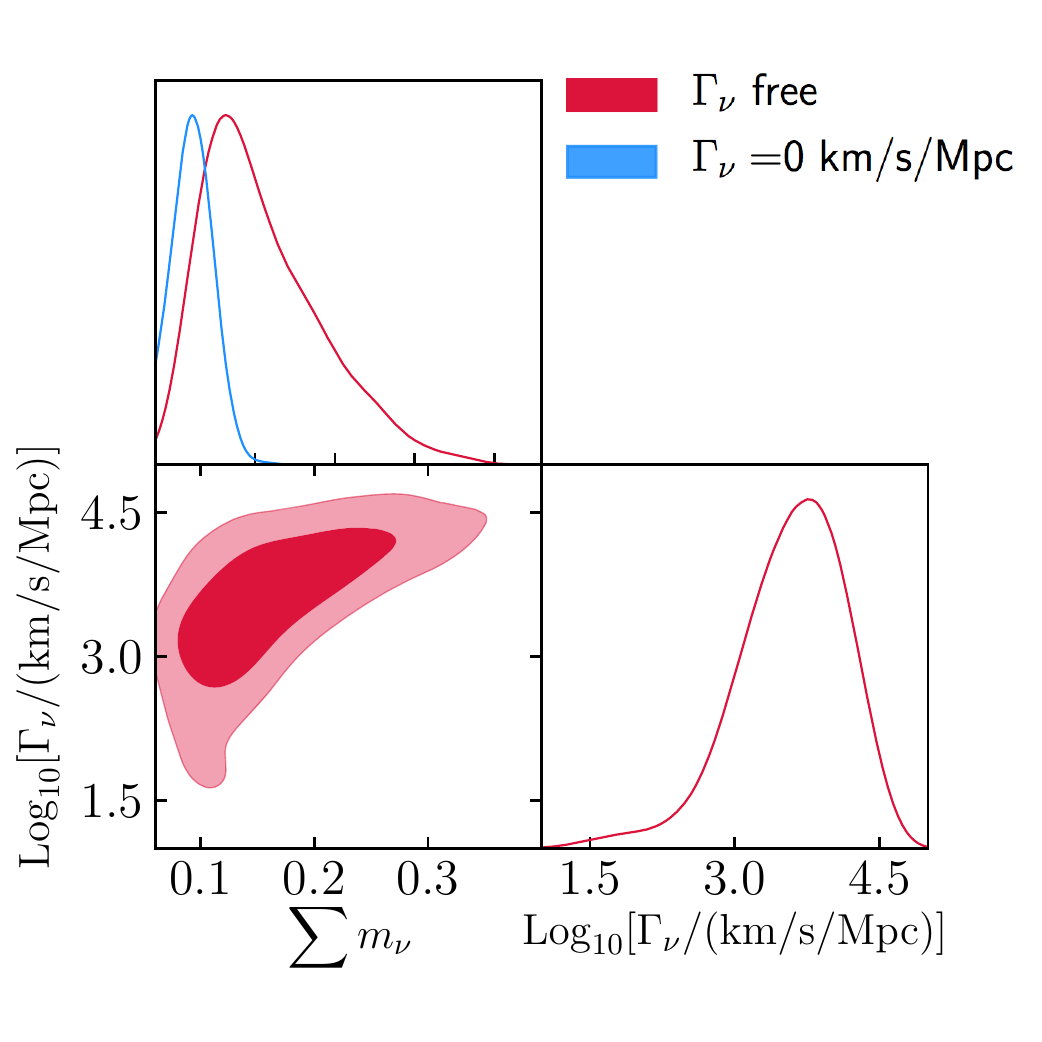}\quad\qquad
    \includegraphics[width=6.5cm]{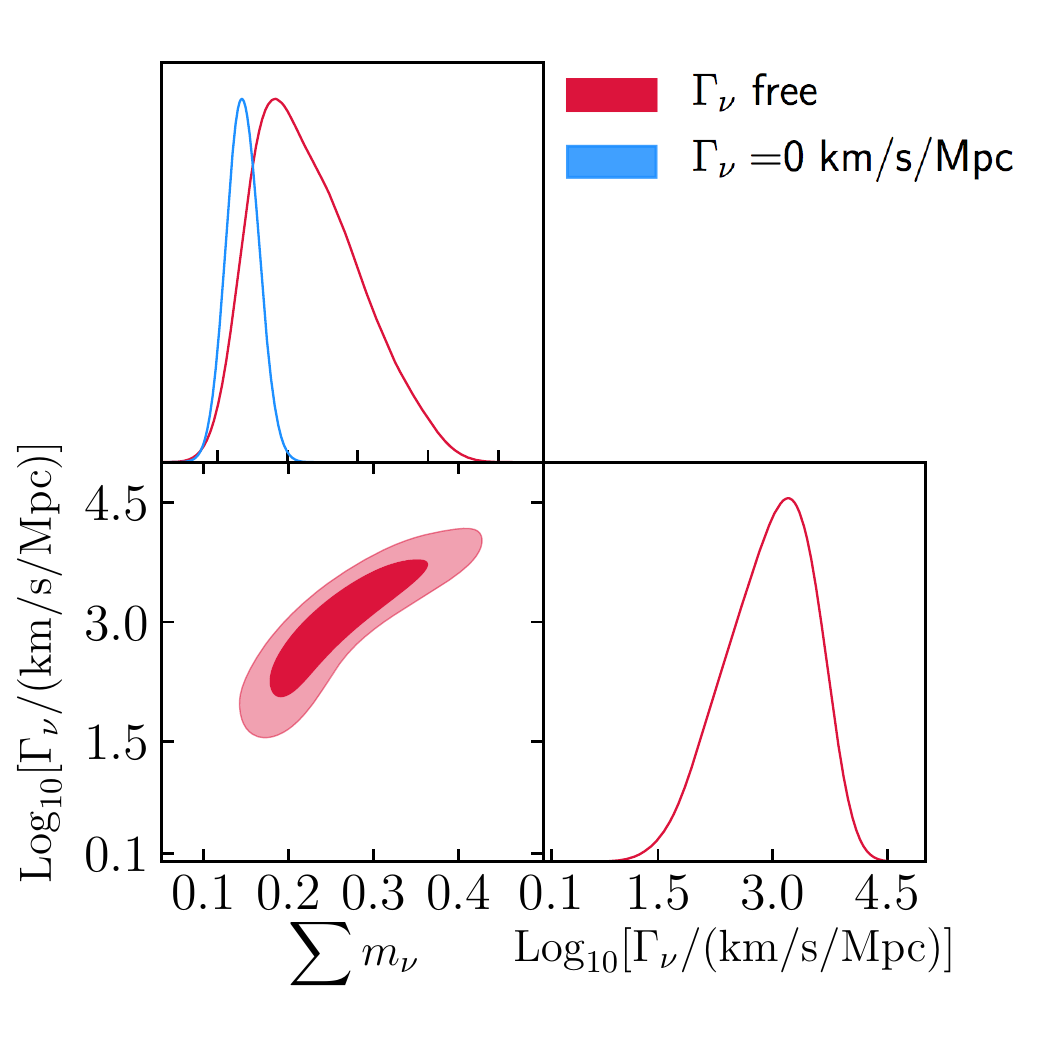}
    \caption{Same as Fig.~\ref{fig:result_stable}, but the fiducial model now assumes decaying neutrinos with $({\rm Log}_{10}[\Gamma_\nu/({\rm km/s/Mpc})], \sum m_\nu/{\rm eV})= 
(3.7,0.16)$ (left panel) and $ (3,0.25)$ (right panel) in the  normal hierarchy.}
    \label{fig:forecasts}
\end{figure*}

\begin{widetext}
\begin{center}
\begin{table}[h]
  \begin{tabular}{|c|c|c|c|c|c|}
    \hline
    \multicolumn{6}{|c|}{ Normal hierarchy} \\
    \hline
    Fiducial $\sum m_\nu/$eV &$0.06$&$0.12$&$0.18$ &$0.24$&$0.30$\\
    \hline
    $\sum m_\nu/$eV& $< 0.085$ &$0.125\pm0.020$  &  $0.183\pm0.017$& $0.243\pm0.016$ & $0.303\pm0.015$\\
    ${\rm Log}_{10}[\Gamma_\nu/{\rm (km/s/Mpc)}]$ & $<3.7$ & $<3.2$ & $<2.1$ & $< 1.7$ & $<1.5$  \\
    \hline
    \multicolumn{6}{|c|}{ Inverted hierarchy} \\
    \hline
     Fiducial $\sum m_\nu/$eV &$0.10$&$0.15$&$0.20$ &$0.25$&$0.30$\\
    \hline
    
    $\sum m_\nu/$eV& $< 0.13$ &$0.154\pm0.017$  &  $0.205^{+0.015}_{-0.017}$& $ 0.253\pm0.016$ & $0.304\pm0.015$\\
    ${\rm Log}_{10}[\Gamma_\nu/{\rm (km/s/Mpc)}]$ & $<2.7$ & $<2.2$ & $<1.8$ & $<1.5$ & $< 1.3$ \\
    \hline
  \end{tabular}
  \caption{Forecast constraints on the sum of neutrino masses (at 68\% C.L.) and decay width of the heaviest neutrino (at 95\% C.L.) from a combination of Planck+Euclid $P(k)$+Euclid Lensing. The fiducial model assumes that neutrinos are stable and that they follow the normal or inverted hierarchy. }
  \label{table:forecast_NH}

  \begin{tabular}{|c|c|c|}
    \hline
    Fiducial $\big({\rm Log}_{10}[\Gamma_\nu/({\rm km/s/Mpc})],\sum m_\nu/{\rm eV}\big)$ &$(3.7,0.16)$ & $(3,0.25)$ \\
    \hline
        $\sum m_\nu/$eV & ~~~~~~~~ $0.167^{+0.035}_{-0.076}$~~~~~~~~& ~~~~~~~~$0.261^{+0.042}_{-0.069}$ ~~~~~~~~ \\
    ${\rm Log}_{10}[\Gamma_\nu/({\rm km/s/Mpc})]$  &  ~~~~~~~~$3.59^{+0.65}_{-0.45}$ ~~~~~~~~& ~~~~~~~~$2.96^{+0.64}_{-0.46}$ ~~~~~~~~ \\
    \hline
      $\sum m_\nu/$eV (stable) & ~~~~~~~~ $0.10\pm0.02$~~~~~~~~& ~~~~~~~~$0.19\pm 0.02$ ~~~~~~~~ \\
	\hline
  \end{tabular}
  \caption{Forecast constraints at 68\% C.L. on the sum of neutrino masses and decay width of the heaviest neutrino from a combination of Planck+Euclid $P(k)$+Euclid Lensing. The models assume a normal neutrino mass hierarchy.}
  \label{table:forecast_decay}
\end{table}
\end{center}

\end{widetext}

Given these constraints on ${\rm Log}_{10}[\Gamma_\nu/{\rm 
(km/s/Mpc)}]$, we anticipate that future cosmological data will be able 
to determine that neutrinos are decaying if the width exceeds this 
limit. To demonstrate this, we turn our attention to a scenario with 
unstable neutrinos and generate two sets of mock data corresponding to 
$({\rm Log}_{10}[\Gamma_\nu/({\rm km/s/Mpc})], \sum m_\nu/{\rm eV})= 
(3.7,0.16)$ and $ (3,0.25)$ with a normal hierarchy. For each mock data 
set and fiducial model we run two cases, one in which we leave 
$\Gamma_\nu$ free to vary and another in which we enforce the constraint 
$\Gamma_\nu = 0$. The purpose of the latter case is to allow us to 
estimate the typical bias that would be introduced if this scenario was 
actually realized in nature and neutrino decays were not accounted for. 

Our results are shown in Fig.~\ref{fig:forecasts} and summarized in 
Table~\ref{table:forecast_decay}. We find, as expected, that for both 
cases the combination of Planck and Euclid sets an upper limit on the 
neutrino lifetime, so that the decaying neutrino scenario can be 
distinguished from the stable case at better than $3\sigma$. Remarkably, 
in both cases we also obtain a lower limit on the neutrino lifetime at 
$3\sigma$, opening the door to the possibility of determining the 
neutrino lifetime from cosmology.

Based on our limits, one might expect that the neutrino lifetime can be 
determined at better than $2\sigma$ provided ${\rm 
Log}_{10}[\Gamma_\nu/({\rm km/s/Mpc})]>3.7$ for $\sum m_\nu/{\rm 
eV}>0.06$.  However, recall that the regime ${\rm 
Log}_{10}[\Gamma_\nu/({\rm km/s/Mpc})] \gtrsim 6$ is not treated in our 
formalism, since neutrinos would be decaying while still relativistic.  
We defer a detailed study of the parameter space for which 
next-generation experiments can determine the neutrino lifetime to 
future work.


Interestingly, we find that in both the cases considered, the precision 
at which $\sum m_\nu$ can be detected is strongly degraded compared to 
the contours in Fig.~\ref{fig:result_stable}. Indeed, in these cases the 
uncertainty on $\sum m_\nu$ is multiplied by $\sim 5$ when $\Gamma$ is 
let free to vary, and $\sim1.5$ when $\Gamma_\nu = 0$ is enforced. This 
is of great importance for next-generation experiments which claim that a 
combination of datasets will be able to detect the sum of neutrino 
masses ``at 5$\sigma$'', even in the minimal mass case. Perhaps even 
more important, we find that when $\Gamma_\nu = 0$ is enforced, a strong 
bias in the reconstructed neutrino mass away from the true value can 
appear. For the specific cases studied here, we find a bias of roughly 
$-0.06$ eV, i.e, a $\sim 3\sigma$ shift away from the ``true'' value.


%


\section{Conclusions}

In summary, we have considered the cosmological signatures of theories 
in which the neutrinos decay into invisible radiation on cosmological 
timescales. We have shown that in this scenario, observations of large 
scale structure made at multiple redshifts may allow two fundamental 
parameters, the sum of neutrino masses and the neutrino lifetime, to be 
determined independently. To assess the prospects for near future 
experiments, we have performed an MCMC analysis based on mock data from 
the Planck and Euclid experiments. Trading Planck data for the mock data 
from a next-generation CMB experiment would strengthen our conclusions. 
In the case of neutrinos that are stable on timescales of order the age 
of the universe, we find that these measurements can improve the lower 
limit on the neutrino lifetime in this scenario by seven orders of 
magnitude, from $\mathcal{O}(10)$ years to 200 million years, without 
significantly impacting the measurement of neutrino mass. In the case of 
neutrinos that decay on timescales shorter than $\mathcal{O}(100)$ 
million years, these measurements may allow the neutrino lifetime to be 
determined from cosmology, provided the neutrinos decay after becoming 
non-relativistic.  We find that in this case, requiring that neutrinos 
are stable when performing the fit can lead to a significant bias in the 
reconstructed neutrino mass. Therefore the possibility that neutrinos 
are unstable on cosmological timescales should be taken into account in 
the analysis of future Euclid data.
 \\\begin{center}
\center{\bf ACKNOWLEDGEMENTS}
\end{center}
\vspace{1em}
 We would like to thank Thejs Brinckmann and Marilena Loverde for useful 
discussions. ZC, AD and YT are supported in part by the National Science 
Foundation under Grant Number PHY-1914731. PD was supported in part by 
NSF grant PHY-1915093.  ZC and YT were also supported in part by the 
US-Israeli BSF grant 2018236. This research used resources of the 
IN2P3/CNRS and the Dark Energy Computing Center funded by the OCEVU 
Labex (ANR-11-LABX-0060) and the Excellence Initiative of Aix-Marseille 
University - A*MIDEX, part of the French Investissements d'Avenir 
programme. YT thanks the Aspen Center for Physics, which is supported by 
National Science Foundation grant PHY-1607611, where part of this work 
was performed.



\bibliography{DecayNu}

\end{document}